\documentclass[conference]{IEEEtran}
\IEEEoverridecommandlockouts
\usepackage{cite}
\usepackage{amsmath,amssymb,amsfonts}
\usepackage{algorithmic}
\usepackage{graphicx}
\usepackage{textcomp}
\usepackage{xcolor}
\usepackage{algorithm,algorithmic}
\def\BibTeX{{\rm B\kern-.05em{\sc i\kern-.025em b}\kern-.08em
    T\kern-.1667em\lower.7ex\hbox{E}\kern-.125emX}}
\begin{document}
	\title{Polarized 6D Movable Antenna for Wireless Communication: Channel Modeling and Optimization}

\author{\IEEEauthorblockN{Xiaodan Shao$\IEEEauthorrefmark{1}$,  Qijun Jiang$\IEEEauthorrefmark{2}$, Derrick Wing Kwan Ng$\IEEEauthorrefmark{3}$, and Naofal Al-Dhahir$\IEEEauthorrefmark{4}$}
	\IEEEauthorblockA{$\IEEEauthorrefmark{1}$Institute for Digital Communications, Friedrich-Alexander-University Erlangen-Nuremberg, Germany\\
		$\IEEEauthorrefmark{2}$School of Science and Engineering, Chinese University of Hong Kong, Shenzhen, China\\
		$\IEEEauthorrefmark{3}$School of Electrical Engineering and Telecommunications, University of New South Wales, Australia\\
		$\IEEEauthorrefmark{4}$Department of Electrical and Computer
		Engineering, University of Texas at Dallas, USA\\
		E-mails: shaoxiaodan@zju.edu.cn, qijunjiang@link.cuhk.edu.cn, w.k.ng@unsw.edu.au, aldhahir@utdallas.edu}\vspace{-32pt}
}\maketitle
\begin{abstract}
	In this paper, we propose a novel polarized six-dimensional movable antenna (P-6DMA) to enhance the performance of wireless communication cost-effectively. Specifically, the P-6DMA enables polarforming by adaptively tuning the antenna's polarization electrically as well as controls the antenna's rotation mechanically, thereby exploiting both polarization and spatial diversity to reconfigure wireless channels for improving communication performance. First, we model the P-6DMA channel in terms of transceiver antenna polarforming vectors and antenna rotations.
	We then propose a new two-timescale
	transmission protocol to maximize the weighted sum-rate for a P-6DMA-enhanced multiuser system. Specifically, antenna rotations at the base station (BS) are first optimized based on the statistical channel state information (CSI) of all users, which varies at a much slower rate compared to their instantaneous CSI. Then, transceiver polarforming vectors are designed to cater to the instantaneous CSI under the optimized BS antennas' rotations.
	Under the polarforming phase shift and amplitude constraints, a new polarforming and rotation joint design problem is efficiently addressed by a low-complexity algorithm based on penalty dual decomposition, where the  polarforming coefficients are updated in parallel to reduce computational time. Simulation results demonstrate the significant performance advantages of polarforming, antenna rotation, and their joint design in comparison with various benchmarks without polarforming or antenna rotation adaptation.
\end{abstract}

\section{Introduction}
As wireless communication systems progress toward the sixth generation (6G) \cite{6gg}, multi-antenna technologies ranging from conventional multiple-input multiple-output (MIMO) to the emerging extra-large-scale MIMO have consistently played a pivotal role in enhancing transmission rates by increasing the number of antennas \cite{larson, wangzhe,shaos, haiquan1}. However, conventional wireless communication systems typically employ fixed-rotation-and-polarization antennas (FRPA), which lack flexibility in dynamically adapting antenna polarization and orientation \cite{haiquan1,duala}. 
When signals propagate through the wireless channel, significant power attenuation inevitably occurs when there is a misalignment between the polarization orientations of the transmitting and receiving antennas \cite{pol1}. 
Moreover, the spatial degrees of freedom
(DoFs) of FRPAs remain limited and cannot efficiently adapt to the dynamic spatial distribution of users in the network \cite{wangzhe}. As each antenna at the transceiver rotates, its polarization state also changes accordingly. Therefore, antenna polarization and rotation are tightly coupled, both of which need to be considered in wireless network designs.

To enable the transceiver's full flexibility in antenna polarization and rotation, we propose a novel polarized six-dimensional movable antenna (P-6DMA), as a new approach to improve the performance of wireless communication without the need to add more antennas and bear their additional cost and energy consumption. 
Specifically, a P-6DMA tunes the polarforming vector with adjustable signal amplitude and phase to control the polarization of each antenna at the transmitter and/or receiver, thus allowing dynamic adjustment of antenna polarization in response to real-time channel conditions and depolarization effects by exploiting polarization diversity. This is achieved through the electronically tunable polarformer, which contains phase shifters and attenuators that collectively adjust the phase and amplitude of the transmitted/received signal (see Fig. \ref{practical_scenario}(a)). 
Furthermore, each P-6DMA can be mechanically rotated independently to reshape the radiation pattern in the angular domain, thereby dynamically adapting to the spatial distribution of the wireless channel and achieving full spatial diversity.  

It is worth to emphasize that the P-6DMA system proposed in this paper differs significantly from the recently-emerged six-dimensional movable antenna (6DMA) \cite{6dmatwc} and traditional polarization reconfigurable antenna \cite{poa}.
Firstly, a mechanically controlled 6DMA, consisting of multiple six-dimensional (6D) rotatable and positionable antennas/subarrays, either lacks polarization adjustment capability or operates with fixed antenna polarization.  Secondly, although traditional polarization reconfigurable antennas can adjust antenna polarization, they only modify the amplitude of the signal while keeping the signal phase and antenna rotation fixed, which does not fully exploit polarization diversity. In stark contrast, the proposed P-6DMA, equipped with a polarformer (see Fig. \ref{practical_scenario}(a)), aims at collectively adjusting both the amplitude and phase of the signals to fully explore the benefits of polarization diversity, as well as adjusting its antenna rotation to enhance spatial DoFs. 

Motivated by the above considerations, we study a P-6DMA-enhanced wireless communication system. In particular, we first propose the innovative P-6DMA architecture and model the corresponding P-6DMA channel. Then, a novel two-timescale optimization problem is proposed to maximize the weighted sum-rate of users under discrete polarforming amplitude and phase shift constraints. In this framework, the slow-timescale antenna rotation at the base station (BS) is optimized leveraging statistical channel state information (CSI) of users, and the fast-timescale discrete transceiver polarforming is designed according to their instantaneous CSI. To handle this problem efficiently, a new penalty dual decomposition (PDD)-based algorithm is proposed, which is shown to achieve higher rates compared to FRPA schemes.

\emph{Notations}: Symbol \((\cdot)^*\) denotes the conjugation operation, \(\mathbf{I}_N\) denotes the \(N \!\times\! N\) identity matrix,  \(\cdot\) and \(\otimes\) denote the dot product and Kronecker product, respectively, \(\|\cdot\|\) and \(\|\cdot\|_\infty\) denote the Euclidean norm and infinity norm of a vector, respectively, and \([\cdot]_j\) denotes the \(j\)-th entry of a vector. 
	\section{System Model} 
\subsection{Polarized 6D Movable Antenna Architecture}
\begin{figure}[t!]
		\setlength{\abovecaptionskip}{-4pt}
	\setlength{\belowcaptionskip}{-15pt}
	\centerline{\includegraphics[width=3.15in]{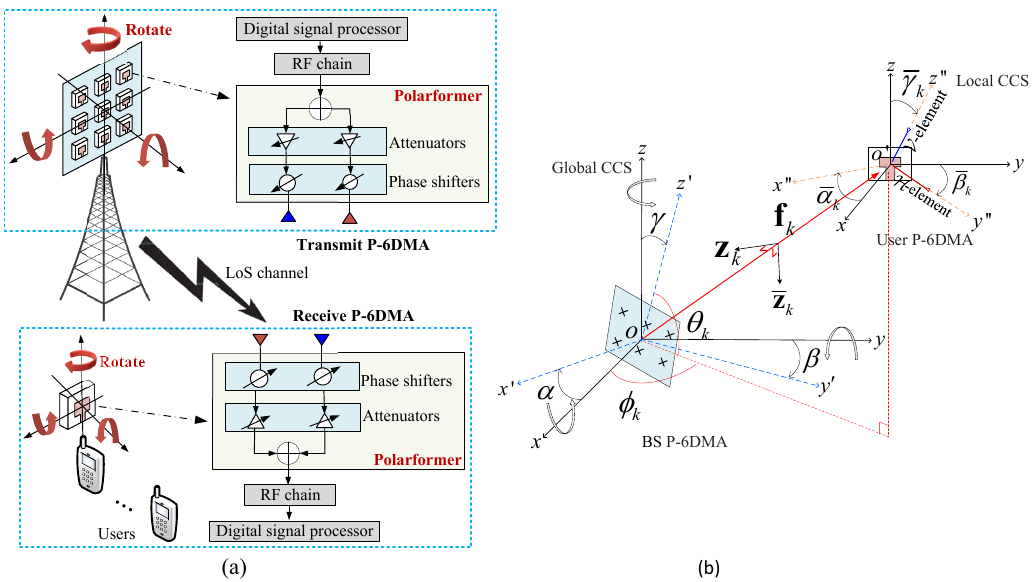}}
	\caption{(a) P-6DMA-enhanced communication system. (b) Geometry illustration of the P-6DMA array at the BS and the P-6DMA at user $k$.}
	\label{practical_scenario}
		\vspace{-0.21cm}
\end{figure}

As illustrated in Fig. \ref{practical_scenario}(a), we consider a downlink communication system, where multiple P-6DMAs are deployed at the BS to serve $K$ single-P-6DMA users. 
Each P-6DMA can independently induce a specific phase shift (via the attached phase shifters) and amplitude change (via the attached attenuators) on the transmit/receive signals, thereby dynamically adjusting the antenna's polarization. The electrically tunable phase shifters and attenuators in the polarformer enable precise polarization control by independently adjusting the phase and amplitude of signals. We assume that each transmit/receive antenna consists of two orthogonally oriented linearly polarized elements, with one element for vertical polarization (\(\mathcal{V}\)-element) and the other for horizontal polarization (\(\mathcal{H}\)-element).
Specifically, the receive polarforming vector for  user \(k\in \{1,2,\cdots,K\}\) is denoted by
\begin{align}
	\mathbf{w}_k = \begin{bmatrix} 
		\rho_{k,1}^{\mathrm{r}}e^{-j \psi_{k,1}^{\mathrm{r}}}, \rho_{k,2}^{\mathrm{r}}e^{-j \psi_{k,2}^{\mathrm{r}}} \end{bmatrix}^T,
\end{align}
where \( \rho_{k,1}^{\mathrm{r}} \in[0,1]\) and \( \rho_{k,2}^{\mathrm{r}}\in[0,1] \) represent the amplitude coefficients for the \(\mathcal{V}\)- and \(\mathcal{H}\)-elements, respectively, of the corresponding antenna at the user. In addition, \( \psi_{k,1}^{\mathrm{r}} \in[0,2\pi)\) and \( \psi_{k,2}^{\mathrm{r}} \in[0,2\pi)\) represent the phase shifts for the user antenna's \(\mathcal{V}\)- and \(\mathcal{H}\)-elements, respectively.

The BS is equipped with \(N > 1\) P-6DMAs, denoted by set \(\mathcal{N} = \{1, 2, \ldots, N\}\). All P-6DMAs at the BS form a uniform planar array (UPA) with a specified size. The antennas within the UPA share identical polarization characteristics determined by the propagation environment. Thus, their polarization can be controlled by the same polarforming vector. Specifically, the transmit polarforming vector at the BS is given by
\begin{align}
	\mathbf{v} = \frac{1}{\sqrt{2}}\begin{bmatrix} 
		\rho_{1}^{\mathrm{t}}e^{-j \psi_{1}^{\mathrm{t}}}, \rho_{2}^{\mathrm{t}}e^{-j \psi_{2}^{\mathrm{t}}} \end{bmatrix}^T,
\end{align}
where \( \rho_{1}^{\mathrm{t}}\! \in \![0,1]\) and \( \rho_{2}^{\mathrm{t}}\in[0,1] \) represent the amplitude coefficients for the \(\mathcal{V}\)- and \(\mathcal{H}\)-elements, respectively, of each antenna at the BS. Similarly, \( \psi_{1}^{\mathrm{t}}\!\in\![0,2\pi)\) and \( \psi_{2}^{\mathrm{t}}\!\in\![0,2\pi)\) represent the phase shifts for the \(\mathcal{V}\)- and \(\mathcal{H}\)-elements, respectively. 

The phase and amplitude of the polarforming vector can be adjusted either continuously or discretely. For ease of practical implementation, we consider the discrete control of amplitude and phase shifts. Specifically, let $Q_{\rho}$ and $Q_{\vartheta}$ denote the number of bits for polarforming amplitude and phase-shift control per antenna element, respectively. We thus have
\begin{align}  
	\!\!\!\![\mathbf{w}_k]_i \text{ and } [\mathbf{v}]_i \in \mathcal{F} \!\triangleq\! \{ \chi \mid  \chi = \rho e^{j \vartheta },  \vartheta \in \mathcal{S},  \rho \in \mathcal{A}\},   
\end{align}  
for $i=\{1,2\}$, where \(\mathcal{F}\) denotes the set of all possible values for \( \chi\), consisting of both amplitude \(\rho\) and phase \(\vartheta\), and 
$\mathcal{S} \triangleq \{0, \frac{2\pi}{D}, \ldots, \frac{2\pi(D-1)}{D}\}$ with $D = 2^{Q_{\vartheta}}$ being the total number of discrete values. Here, the discrete phase values are assumed to be equally spaced in the interval $[0, 2\pi)$, \(\mathcal{A} \triangleq \{{\rho}_1, \ldots, {\rho}_{2^{Q_{
			\rho}}}\}\) denotes the controllable amplitude set with \(|\mathcal{A}| = 2^{Q_{
		\rho}}\), and \(\rho_i\)'s are equally spaced in the interval \([0, 1]\) for $i \in \{1, \ldots, 2^{Q_\rho}\}$. 

In this paper, we assume that the orientation of each user's P-6DMA is arbitrary due to random rotation, while the P-6DMA array at the BS can be actively controlled to rotate freely about its center. This rotation can be realized through mechanical control, such as electric motors and precision gears, or by manual adjustment to achieve physical movement. To facilitate the description of the user/BS's antenna rotation, we establish three distinct Cartesian coordinate systems (CCSs). As shown in Fig. \ref{practical_scenario}(b), the global CCS is denoted by \(o\text{-}xyz\), with the BS center at the origin \(o\), the BS P-6DMA array’s local CCS is denoted by \(o'\text{-}x'y'z'\), with the array center as the origin \(o'\), and each user's local CCS is denoted by \(o''\text{-}x''y''z''\), with the center of the user's antenna as origin \(o''\).

On the BS side, the rotation of the P-6DMA array can be characterized by the following rotation angle vector
\begin{align}\label{bb}
	\mathbf{u}=[\alpha,\beta,\gamma]^T\in \mathbb{R}^{3 \times 1},
\end{align}
where $\alpha$,  $\beta$, and $\gamma$ all in \([0, 2\pi)\)  denote the rotation angles of the P-6DMA array at the BS with respect to (w.r.t.) the $x$-axis, $y$-axis, and $z$-axis in the global CCS, respectively.
Given $\mathbf{u}$, the corresponding rotation matrix can be written as
\begin{align}\label{R}
	&\!\!\!\mathbf{R}(\mathbf{u})
	&\!=\!\begin{bmatrix}
		c_{\beta}c_{\gamma} & c_{\beta}s_{\gamma} & -s_{\beta} \\
		s_{\beta}s_{\alpha}c_{\gamma}-c_{\alpha}s_{\gamma} & s_{\beta}s_{\alpha}s_{\gamma}+c_{\alpha}c_{\gamma} & c_{\beta}s_{\alpha} \\
		c_{\alpha}s_{\beta}c_{\gamma}+s_{\alpha}s_{\gamma} & c_{\alpha}s_{\beta}s_{\gamma}-s_{\alpha}c_{\gamma} &c_{\alpha}c_{\beta} \\
	\end{bmatrix},\!\!
\end{align}
where $c_{x}=\cos(x)$ and $s_{x}=\sin(x)$ \cite{6dmatwc}. Let \(\bar{\mathbf{r}}_{n}\) denote the position of the \(n\)-th antenna of the P-6DMA array at the BS in its local CCS. Then, the position of the \(n\)-th antenna at the BS in the global CCS is given by $
\mathbf{r}_{n}(\mathbf{u})=\mathbf{R}
(\mathbf{u})\bar{\mathbf{r}}_{n}$.
Next, on each user side, let 
\begin{align}\label{URR}
	\mathbf{u}_{k}^{\mathrm{r}}=[\bar{\alpha}_k,\bar{\beta}_k,\bar{\gamma}_k]^T\in \mathbb{R}^{3 \times 1},
\end{align}
denote the rotation angle vector of the $k$-th user's local CCS \(o''\text{-}x''y''z''\) relative to \(o\text{-}xyz\), where \(\bar{\alpha}_k\), \(\bar{\beta}_k\), and \(\bar{\gamma}_k\) all in \([0, 2\pi)\) denote the rotation angles of the $k$-th user w.r.t. the $x$-axis, $y$-axis and $z$-axis in the global CCS, respectively (see Fig. \ref{practical_scenario}(b)). Similar to \eqref{R}, given \(\mathbf{u}_{k}^{\mathrm{r}}\), the corresponding rotation matrix is denoted by \(\mathbf{R}(\mathbf{u}_{k}^{\mathrm{r}})\), which is omitted for brevity.
%\vspace{-3pt}	
\subsection{Polarized 6D Movable Antenna Channel Model}
%\vspace{-3pt}
For the purpose of exposition, we assume that the channel between the BS and each user is a far-field LoS channel, and each P-6DMA at the user is an omnidirectional antenna. Let $\phi_k\!\in\![-\pi,\pi]$ and $\theta_k\!\in\![-\pi/2,\pi/2]$ denote the azimuth and elevation angles, respectively, of the signal from user $k$ arriving at the BS w.r.t. its center.
The corresponding pointing vector in direction $(\theta_k, \phi_k)$ is thus given by
$		\mathbf{f}_k\!=\![\cos(\theta_k)\cos(\phi_k),\! \cos(\theta_k)\sin(\phi_k),\!\sin(\theta_k)]^T$.
Consequently, the steering vector of the P-6DMA array at the BS is given by
\begin{align}\label{gen}
	\mathbf{a}_{k}(\mathbf{u})\!=\! \left[e^{-j\frac{2\pi}{\lambda}
		\mathbf{f}_k^T\mathbf{r}_{1}(\mathbf{u})},
	\!\cdots,\! e^{-j\frac{2\pi}{\lambda}\mathbf{f}_k^T
		\mathbf{r}_{N}(\mathbf{u})}\right]^T,
\end{align}
where $\lambda$ denotes the signal carrier wavelength.
Next, to determine effective antenna gain $g_{k}(\mathbf{u})$, we project \( \mathbf{f}_k \) onto the local CCS of the P-6DMA array, denoted by $
\tilde{\mathbf{f}}_{k}=\mathbf{R}(\mathbf{u})^{-1}\mathbf{f}_k$.	Then, we represent $	\tilde{\mathbf{f}}_{k}$ in the spherical coordinate system as
$	\tilde{\mathbf{f}}_{k}\!=\![\cos(\tilde{\theta}_{k})\cos(\tilde{\phi}_{k}), \cos(\tilde{\theta}_{k})\sin(\tilde{\phi}_{k}), \sin(\tilde{\theta}_{k})]^T$, where $\tilde{\theta}_{k}$ and $\tilde{\phi}_{k}$ represent the directions of arrival (DoAs) in the local CCS.
Finally, the effective antenna gain $g_{k}(\mathbf{u})$ of the P-6DMA array at BS along direction \((\tilde{\theta}_{k}, \tilde{\phi}_{k})\) in linear scale is given by
$g_{k}(\mathbf{u})\!=\!10^{\frac{G(\tilde{\theta}_{k}, 	\tilde{\phi}_{k})}{10}}$,
where \(G(\tilde{\theta}_{k}, \tilde{\phi}_{k})\) denotes the effective antenna gain in dBi determined by the antenna's radiation pattern \cite{6dmatwc}. Thus, the {\emph{unpolarformed channel}} between the P-6DMA array at the BS and user \(k\) is given by 
\begin{align}\label{uik}
	\mathbf{h}_{k}^{\mathrm{LoS}}(\mathbf{u})=		\sqrt{\nu_k}e^{-j\frac{2\pi d_k}{\lambda}}\sqrt{g_{k}(\mathbf{u})}
	\mathbf{a}_{k}(\mathbf{u})\in \mathbb{C}^{N\times 1},
\end{align}
where \(\nu_k \in \mathbb{R}^1\) denotes the path loss coefficient and $d_k$ denotes the distance between the $k$-th user's location and the BS center.

As shown in Fig. \ref{practical_scenario}(b), the \(\mathcal{V}\)-element of the antenna is aligned along the positive \(y'\)- or \(y''\)-axis, while the \(\mathcal{H}\)-element is oriented along the positive \(x'\)- or \(x''\)-axis, with their unit vectors given by \(\mathbf{e}_\mathrm{v} = [0,1,0]^T\) and \(\mathbf{e}_\mathrm{h} = [1,0,0]^T\), respectively. Moreover, the polarization state of an electromagnetic (EM) wave can be described by any two orthogonal electric field components on the wavefront, given by \(\mathbf{z}_k = [s_{\theta_k}s_{\phi_k}, -c_{\theta_k}, s_{\theta_k}c_{\phi_k}]^T\) and \(\bar{\mathbf{z}}_k = [c_{\phi_k}, 0, -s_{\phi_k}]^T\) \cite{heap}.
Consequently, the transmit field components of the LoS path are generated by projecting the transmit antenna's electric fields onto the LoS signal direction. The corresponding transformation is  given by
\begin{align}
	\!\!\!\!\!	\mathbf{P}_{k}(\mathbf{u}) \!= \!
	\begin{bmatrix}
		(\mathbf{R}(\mathbf{u})\mathbf{e}_\mathrm{v}) \cdot \mathbf{z}_k & (\mathbf{R}(\mathbf{u})\mathbf{e}_\mathrm{h}) \cdot \mathbf{z}_k \\
		(\mathbf{R}(\mathbf{u})\mathbf{e}_\mathrm{v}) \cdot \bar{\mathbf{z}}_k & (\mathbf{R}(\mathbf{u})\mathbf{e}_\mathrm{h}) \cdot \bar{\mathbf{z}}_k
	\end{bmatrix}\in \mathbb{C}^{2\times 2},
\end{align}
where $\mathbf{R}(\mathbf{u})\mathbf{e}_\mathrm{v}$ and $\mathbf{R}(\mathbf{u})\mathbf{e}_\mathrm{h}$ represent the mapping of the local CCS of the \(\mathcal{V}\)-element/\(\mathcal{H}\)-element of the P-6DMA array at the BS to the global CCS.
Similarly, the receive field components are obtained by projecting the LoS signal direction onto the receive antenna exploiting the projection matrix
\begin{align}
	\!\!\!\!\!\mathbf{Q}_k (\mathbf{u}_{k}^{\mathrm{r}})\!= \!
	\begin{bmatrix}
		\mathbf{z}_k \cdot (\mathbf{R}(\mathbf{u}_{k}^{\mathrm{r}})\mathbf{e}_\mathrm{v}) & \bar{\mathbf{z}}_k \cdot (\mathbf{R}(\mathbf{u}_{k}^{\mathrm{r}})\mathbf{e}_\mathrm{v}) \\
		\mathbf{z}_k \cdot (\mathbf{R}(\mathbf{u}_{k}^{\mathrm{r}})\mathbf{e}_\mathrm{h}) & \bar{\mathbf{z}}_k \cdot (\mathbf{R}(\mathbf{u}_{k}^{\mathrm{r}})\mathbf{e}_\mathrm{h})
	\end{bmatrix}\in \mathbb{C}^{2\times 2},
\end{align}
where $\mathbf{R}(\mathbf{u}_{k}^{\mathrm{r}})\mathbf{e}_\mathrm{v}$ and $\mathbf{R}(\mathbf{u}_{k}^{\mathrm{r}})\mathbf{e}_\mathrm{h}$ denote the transformation of the local CCS of the \(\mathcal{V}\)-element/\(\mathcal{H}\)-element of the user's P-6DMA to the global CCS.
Consequently, the dual-polarized response matrix between the $k$-th user and the dual-polarized antennas on the P-6DMA array at the BS is given by  
\begin{align}\label{po3}
	{\mathbf{A}_{k}(\mathbf{u},\mathbf{u}_{k}^{\mathrm{r}})} = \mathbf{Q}_k (\mathbf{u}_{k}^{\mathrm{r}})\mathbf{P}_{k}(\mathbf{u})\in \mathbb{C}^{2\times 2}.
\end{align}

%Since the phase variation due to the LoS channel is identical for both the \(\mathcal{V}\)- and \(\mathcal{H}\)-ports of the antenna regardless of the transmit-receive pair, 
Next, the LoS channel from the \(\mathcal{V}\)- and \(\mathcal{H}\)-ports of user \(k\) to those of the P-6DMA array at the BS is expressed as
\begin{align}\label{tu5}
	\!\! \overline{\mathbf{h}}_{k}(\mathbf{u}) = \mathbf{h}_{k}^{\mathrm{LoS}}(\mathbf{u}) \otimes \mathbf{A}_{k}(\mathbf{u},\mathbf{u}_{k}^{\mathrm{r}})\in \mathbb{C}^{2N\times 2},
\end{align}
which captures the wireless signal propagation characteristics across all potential polarization states.
Different from dual-polarized antennas \cite{duala}, which require two radio frequency (RF) chains per antenna (one for each port), each proposed P-6DMA is connected to a single RF chain for subsequent signal processing, thereby reducing hardware cost and circuit power consumption (see Fig. \ref{practical_scenario}(a)). The overall channel, obtained by applying the receive polarforming vector \(\mathbf{w}_k\) and transmit polarforming vector \(\mathbf{v}\) to \eqref{tu5}, is therefore given by
\begin{subequations}
\begin{align}
	&\mathbf{h}_{k}(\mathbf{u},\mathbf{w}_k,\mathbf{v}) =(\mathbf{I}_N\otimes\mathbf{v}^H)\overline{\mathbf{h}}_{k}
	(\mathbf{u})\mathbf{w}_k
	\label{pcc}\\
	&=\underbrace{\mathbf{h}_{k}^{\mathrm{LoS}}(\mathbf{u})}_{\text{Unpolarformed channel}}\times~ 
	\underbrace{\left(\mathbf{v}^H\mathbf{A}_{k}(\mathbf{u},\mathbf{u}_{k}^{\mathrm{r}})\mathbf{w}_k\right)}_{\text{Polarformed channel}}\in \mathbb{C}^{N\times 1}.\label{pcc1}
\end{align}
\end{subequations}

\textbf{Remark 1:} In \eqref{pcc1}, the unpolarformed channel remains unaffected when varying the polarforming vectors and user rotations. This distinctive P-6DMA-specific channel representation can be leveraged to design efficient channel estimation algorithms by exploiting the common parameters embedded in the unpolarformed channel across different controllable polarforming vectors, which is left for future work. Furthermore, the elegant structure of \eqref{pcc1} facilitates the optimization of polarforming at the transceivers (see Section III).

\vspace{-3pt}
\subsection{Transmission Protocol}
\vspace{-3pt}
In the proposed P-6DMA-enhanced communication system, the rotations of antennas are mechanically controlled, which inherently limits their adjustment speed, rendering them more suitable for adapting to slow-timescale channel variation, i.e., over a
long period of time during which the statistical CSI remains approximately constant. In contrast, the transmit and receive polarforming of P-6DMA are electronically controlled, thus allowing their rapid adjustments to adapt to fast-timescale channel variations. Building upon this insight, we propose a hierarchical two-timescale transmission scheme. Specifically, the considered time frame \(T\) is divided into two phases: a slow-timescale phase \(T_\mathrm{s}\) for tuning antenna rotation at the BS based on statistical CSI of users, and a fast-timescale phase \(T_\mathrm{c}\) for adapting polarforming to their instantaneous CSI.
 \vspace{-3pt}
\section{Polarforming and Rotation Optimization}
\vspace{-2pt}
As shown in Fig. \ref{practical_scenario}(a), the downlink received signal at user $k$ is given by \(y_k = \mathbf{h}_k^H(\mathbf{u},\mathbf{w}_k,\mathbf{v}) \sum_{j=1}^{K}\mathbf{c}_j x_j + n_k\), where \(x_k \!\sim\! \mathcal{CN}(0, 1)\) denotes the complex information symbol with zero mean and unit variance, \(\mathbf{c}_k \!\in\! \mathbb{C}^{N\times 1}\) is its transmit digital precoder, and \(n_k \!\sim\! \mathcal{CN}(0,\sigma^2)\) is noise with variance $\sigma^2$. Defining  \(\mathbf{c}=\{\mathbf{c}_k\}_{k=1}^{K}\), the achievable rate for user \(k\) over the channel is \(R_k(\mathbf{u},\mathbf{w}_k,\mathbf{v},\mathbf{c}) \!=\! \log_2(1\!+\!\frac{|\mathbf{h}_k^H(\mathbf{u},\mathbf{w}_k,\mathbf{v}) \mathbf{c}_k|^2}{\sum_{j\neq k} |\mathbf{h}_k^H(\mathbf{u},\mathbf{w}_k,\mathbf{v}) \mathbf{c}_j|^2+\sigma^2})\).

Following the proposed two-timescale protocol,
we aim to maximize the weighted sum-rate for all users
by jointly optimizing BS-side antenna rotations $\mathbf{u}$ in the slow
timescale, as well as BS-side transmit polarforming vectors $\mathbf{v}$ and precoding vectors $\mathbf{c}$ and user-side polarforming vectors $\{\mathbf{w}_k\}$ in the fast timescale. This leads to the following optimization problem: 
\begin{subequations}
	\label{MG3}
	\begin{align}
		\!\text{(P1) :}~\!\!\mathop{\max}\limits_{\mathbf{u}}~&~\mathbb{E}\bigl[
		\mathop{\max}\limits_{\{\mathbf{w}_k\}_{k=1}^K, \mathbf{v},\mathbf{c}}
		\sum\limits_{k\in\mathcal{K}}\varrho_k{R}_k(
		\mathbf{u},\mathbf{w}_k,\mathbf{v},\mathbf{c})\bigr]\\
		\text {s.t.}~&~[\mathbf{w}_k]_i\in \mathcal{F}, \forall k\in \mathcal{K}, i\in \{1,2\}, \label{pc1}\\
		~&~ [\mathbf{v}]_i\in \mathcal{F}, i\in \{1,2\}, \label{pc2}\\
		~&~\sum_{k \in \mathcal{K}} \|\mathbf{c}_k\|^2 \leq \zeta, \label{pow1}\\
		~&~[\mathbf{u}]_i\in [0,2\pi], ~\forall i \in \{1,2,3\},  \label{M1}
	\end{align}
\end{subequations}
where $\varrho_k\geq 0$ represents the rate weight of user $k$, $\zeta$ is the total transmit power of the BS, and the expectation in the objective function is taken over random channel variations due to arbitrary user locations and rotations. Also, constraints \eqref{pc1} and \eqref{pc2} ensure that the receive and transmit polarforming vectors satisfy the discrete amplitude and phase requirements, respectively. Constraint \eqref{M1} guarantees that each rotation angle of the P-6DMA array remains within the range $[0,2\pi]$.
	\vspace{-3pt}
\subsection{Polarforming Optimization}
	\vspace{-3pt}
During each channel coherence interval, the BS first estimates the instantaneous channels of all users,  $\mathbf{h}_k(\mathbf{u},\mathbf{w}_k,\mathbf{v})$, for all possible $\{\mathbf{w}_k\}=\{\mathbf{w}_k\}_{k=1}^{K}$ and $\mathbf{v}$ and with fixed antenna rotation vector $\mathbf{u}$. Then, the BS determines its polarforming $\mathbf{v}$, transmit precoding $\mathbf{c}$ as well as user polarforming $\{\mathbf{w}_k\}$. Given $\mathbf{u}$, problem (P1) is simplified to 
\begin{subequations}
	\label{op2}
	\begin{align}
		\text{(P2) :}~&~
		\mathop{\max}\limits_{\{\mathbf{w}_k\}_{k=1}^K, \mathbf{v}, \mathbf{c}}
		\sum\limits_{k\in\mathcal{K}}\varrho_k{R}_k(
		\mathbf{u},\mathbf{w}_k,\mathbf{v},\mathbf{c})\\
		~&~\text {s.t.}~\eqref{pc1},\eqref{pc2},\eqref{pow1}.
	\end{align}
\end{subequations}

To reformulate problem (P2) into a more tractable form, we employ the weighted minimum mean
squared error (WMMSE) method \cite{38}.
The mean square error (MSE) for user \( k \), defined as \( e_k = \mathbb{E}[ |\xi_k y_k - x_k|^2 ] \), is derived as  
$e_k = |\xi_k|^2 ( \sum_{j\in\mathcal{K} }|\mathbf{h}_k(\mathbf{u},\mathbf{w}_k,\mathbf{v})^H
\mathbf{c}_j|^2+\sigma^2 )
- 2 \operatorname{Re} \{ \xi_k^* \mathbf{h}_k(\mathbf{u},\mathbf{w}_k,\mathbf{v})^H
\mathbf{c}_k \} + 1$, where \( \xi_k \) denotes the equalizer coefficient.
Then, problem (P2) shares the same globally optimal solution with the following WMMSE problem:
\begin{subequations}
	\label{op2-1}
	\begin{align}
		\text{(P2-1) :}~&~
		\mathop{\min}\limits_{\{\mathbf{w}_k,\xi_k,\epsilon_k\}_{k=1}^K, \mathbf{v},\mathbf{c}}~~\sum\limits_{k\in\mathcal{K}}\varrho_k(\epsilon_ke_k-\log_2(\epsilon_k))
		\\
		~&~\text {s.t.}~\eqref{pc1},\eqref{pc2}, \eqref{pow1},
	\end{align}
\end{subequations}
where $\epsilon_k$ denotes the weighting factor for user $k$.
To facilitate parallel and element-wise updates of the elements in polarforming vectors $\{\mathbf{w}_k\}$ and $\mathbf{v}$, thereby simplifying their optimization,
we introduce auxiliary optimization variables $\{\overline{\mathbf{w}}_k\}$ and $\{\overline{\mathbf{v}}\}$. As a result, problem (P2-1) is equivalently transformed to
\begin{subequations}
	\label{aop2-1}
	\begin{align}
		\text{(P2-2) :}&
		\mathop{\min}\limits_{\{\mathbf{w}_k,\xi_k,\epsilon_k\}_{k=1}^K, \mathbf{v}, \mathbf{c}}~\sum\limits_{k\in\mathcal{K}}\varrho_k(\epsilon_ke_k-\log_2(\epsilon_k))
		\\
		\text {s.t.}&~\sum_{k \in \mathcal{K}} \|\mathbf{c}_k\|^2 \leq \zeta, \label{3pow1}\\
		~&~\mathbf{w}_k=\overline{\mathbf{w}}_k, k\in\mathcal{K}, \label{0pcc1}\\
		~&~\mathbf{v}=\overline{\mathbf{v}},\label{0pcc01}\\
		~&~[\overline{\mathbf{w}}_k]_i\in \mathcal{F}, \forall k\in \mathcal{K}, i\in \{1,2\}, \label{3pc1}\\
		~&~ [\overline{\mathbf{v}}]_i\in \mathcal{F}, i\in \{1,2\}. \label{3pc2}
	\end{align}
\end{subequations}
Next, we employ the PDD to develop a nested-loop iterative algorithm with inner and outer loops for solving (P2-2). 
\subsubsection{Inner-loop for addressing (P2-2)}
Specifically, in the inner loop of PDD, we apply the block coordinate descent (BCD) method to address the following augmented Lagrangian problem of (P2-2):
\begin{subequations}
	\label{aop33-1}
	\begin{align}
		\!\!\!\!\text{(P2-3) :}&
		\mathop{\min}\limits_{\{\mathbf{w}_k,\xi_k,\epsilon_k\}_{k=1}^K, \mathbf{v},\mathbf{c}}~~\sum\limits_{k\in\mathcal{K}}\varrho_k(\epsilon_ke_k-\log_2(\epsilon_k))+\nonumber\\
		&\!\!\!\frac{1}{2 \mu }\!\sum_{k\in\mathcal{K}}\|\mathbf{w}_k-\overline{\mathbf{w}}_k+ \mu\mathbf{t}_k \|^2	\!+\!\frac{1}{2\mu} \!\left\| \mathbf{v} \!-\! \overline{\mathbf{v}} \!+\! \mu \bar{\mathbf{t}} \right\|^2,\label{hl9}
		\\
		\text {s.t.}&~\eqref{3pow1}, \eqref{3pc1},\eqref{3pc2}, 
	\end{align}
\end{subequations}
where \(\mathbf{t}_k\) and \(\bar{\mathbf{t}}\) represent the dual variables corresponding to  constraints \(\mathbf{w}_k\!=\!\overline{\mathbf{w}}_k\) and \(\mathbf{v}\!=\!\overline{\mathbf{v}}\), respectively, and \(\mu\) is the penalty factor. Dividing the variables into blocks \(\{\mathbf{w}_k\}\), \(\{\bar{\mathbf{w}}_k\}\), \(\{\mathbf{v}\}\), \(\{\bar{\mathbf{v}}\}\), \(\{\xi_k\}\), \(\{\epsilon_k\}\), and \(\{\mathbf{c}_k\}\) allows each block to be optimized independently while holding the others fixed.

First, the user polarforming vectors \(\{\mathbf{w}_k\}\) are updated by solving the following unconstrained quadratic program (QP) problem:
\label{op25}
\begin{align}
	\!\text{(P2-3.1) :}~&
	\mathop{\min}\limits_{\{\mathbf{w}_k\}}~~\sum\limits_{k\in\mathcal{K}}\varrho_k\epsilon_k |\xi_k|^2 
	\sum_{j\in\mathcal{K}}
	\bigl|
	\mathbf{w}_k^H
	\mathbf{M}_{k}^H
	\mathbf{c}_j
	\bigr|^2
	-\sum_{k\in\mathcal{K}}2\varrho_k\epsilon_k\nonumber\\
	&\!\!\!\!\!\!\!\!\!\!\!\!\!\!\!
	\mathrm{Re}\Bigl\{
	\xi_k^*
	\mathbf{w}_k^H
	\mathbf{M}_{k}^H
	\mathbf{c}_k
	\Bigr\}+\frac{1}{2 \mu }\sum_{k\in\mathcal{K}}\|\mathbf{w}_k-\overline{\mathbf{w}}_k+ \mu\mathbf{t}_k \|^2,
\end{align}
which is derived by substituting the relationship $\mathbf{h}_k( \mathbf{u}, \mathbf{w}_k, \mathbf{v})
=
\mathbf{M}_k \mathbf{w}_k$ into problem (P2-3) and neglecting the terms that do not involve 
$\{\mathbf{w}_k\}$, where
$\mathbf{M}_k =
\mathbf{h}_{k}^{\mathrm{LoS}} \mathbf{v}^H \mathbf{A}_{k}
\in \mathbb{C}^{N \times 2}$.
The closed-form optimal solution of (P2-3.1) can be expressed as
\begin{align}\label{50}
	\mathbf{w}_k^{\mathrm{opt}}
	=
	\mathbf{C}_k^{-1}\mathbf{b}_k,
\end{align}
where $	\mathbf{C}_k 
=
2\varrho_k\epsilon_k|\xi_k|^2\sum_{j\in\mathcal{K}} \mathbf{M}_k^H\mathbf{c}_j\mathbf{c}_j^H\mathbf{M}_k 
+ \tfrac{1}{\mu}\mathbf{I}$ and $
\mathbf{b}_k = 
2\varrho_k\epsilon_k\xi_k\mathbf{M}_k^H\mathbf{c}_k
+ \tfrac{1}{\mu}(\overline{\mathbf{w}}_k - \mu\mathbf{t}_k)$.

Subsequently, the $\overline{\mathbf{w}}_k$-subproblem is
given by  
\begin{subequations}
	\label{ppp}
	\begin{align}
		\text{(P2-3.2) :} ~\min_{\overline{\mathbf{w}}_k}~&~\|\mathbf{w}_k - \overline{\mathbf{w}}_k + \mu \mathbf{t}_k \|^2  \\
		\text {s.t.}~&~[\overline{\mathbf{w}}_k]_i \in \mathcal{F}, \quad \forall i \in \{1, 2\}.
	\end{align}
\end{subequations}
Since the elements of \(\overline{\mathbf{w}}_k\) are independent in both the objective function and constraints, the optimal solution can be computed in parallel as
\begin{align}\label{54}
	[\overline{\mathbf{w}}_k]_i^{\text{opt}} = \hat{\rho}_{k,i} e^{j \angle [\overline{\mathbf{w}}_k]_i},
\end{align} 
where $	\angle [\overline{\mathbf{w}}_k]_i = \arg \min_{\angle [\overline{\mathbf{w}}_k]_i \in \mathcal{S}} |\angle [\overline{\mathbf{w}}_k]_i - \angle ([\mathbf{w}_k]_i+\mu [\mathbf{t}_k]_i )|$ and $
\hat{\rho}_{k,i} = \arg \min_{{\rho}_{k,i} \in \mathcal{A}} |{\rho}_{k,i} e^{j \angle [\overline{\mathbf{w}}_k]_i} - ([\mathbf{w}_k]_i+\mu [\mathbf{t}_k]_i )|$.

Then, the update of the BS polarforming vectors $\mathbf{v}$ can be conducted by solving the following unconstrained QP problem:
\begin{align}	\label{hv1}
	&\!\!\text{(P2-3.3) :}~
	\mathop{\min}\limits_{\mathbf{v}}~~\sum\limits_{k\in\mathcal{K}}\varrho_k\epsilon_k|\xi_k|^2 \sum_{j\in\mathcal{K} }|{\varepsilon}_{k,j}\mathbf{v}^H\widehat{\mathbf{m}}_{k}|^2\nonumber\\
	&\!\!-\sum\limits_{k\in\mathcal{K}}\varrho_k\epsilon_k2 \operatorname{Re} \big\{ \xi_k^* \varepsilon_{k,k}\mathbf{v}^H\widehat{\mathbf{m}}_{k}\big\}
	+\frac{1}{2\mu} \left\| \mathbf{v} - \overline{\mathbf{v}} + \mu \bar{\mathbf{t}} \right\|^2,
\end{align}
where $\widehat{\mathbf{m}}_{k}=\mathbf{A}_{k}\mathbf{w}_k$ and $\varepsilon_{k,j}=(\mathbf{h}_{k}^{\mathrm{LoS}})^T\mathbf{c}_{j}$.
The optimal solution of (P2-3.3) can be expressed as
\begin{align}
	\mathbf{v}^{\mathrm{opt}} = \bar{\mathbf{C}}^{-1} \bar{\mathbf{\mathbf{b}}},
\end{align}
where
$\bar{\mathbf{C}}\! =\! \sum_{k \in \mathcal{K}} \varrho_k \epsilon_k |\xi_k|^2\sum_{j \in \mathcal{K}}2 \varepsilon_{k,j} \varepsilon_{k,j}^* \widehat{\mathbf{m}}_{k} \widehat{\mathbf{m}}_{k}^H \!+\! \frac{1}{\mu} \mathbf{I}$ and $
\bar{\mathbf{b}}\! = \!\sum_{k \in \mathcal{K}} \varrho_k \epsilon_k 2 \xi_k^* \varepsilon_{k,k} \widehat{\mathbf{m}}_{k} + \frac{1}{\mu} \left( \overline{\mathbf{v}} - \mu \bar{\mathbf{t}} \right)$.

	Next, the $\overline{\mathbf{v}}$-subproblem is given by 
	\begin{subequations}
		\label{ovv}
		\begin{align}
			\text{(P2-3.4) :}~\min_{\overline{\mathbf{v}}}~&~\|\mathbf{v}-\overline{\mathbf{v}}  + \mu \bar{\mathbf{t}}\|^2  \\
			\text {s.t.}~&~[\overline{\mathbf{v}}]_i \in \mathcal{F}, \quad \forall i \in \{1, 2\}.
		\end{align}
	\end{subequations}
	Similar to problem (P2-3.2), the optimal solution of problem (P2-3.4) can be efficiently computed in parallel, and its expression is omitted for brevity. Next, minimizing $\sum_{k \in \mathcal{K}} \varrho_k\epsilon_ke_k$ leads to the linear minimum mean square error (LMMSE) equalizer coefficient $
	\xi_k^{\mathrm{opt}}=\frac{\mathbf{h}_k(\mathbf{u},\mathbf{w}_k,\mathbf{v})^H
		\mathbf{c}_k}{\sum_{j\in\mathcal{K} }|\mathbf{h}_k(\mathbf{u},\mathbf{w}_k,\mathbf{v})^H
		\mathbf{c}_j|^2+\sigma^2}$, and the optimal solution for $\epsilon_k$
	is  $\epsilon_k^{\mathrm{opt}} = \frac{1}{e_k}$. The transmit precoder update is obtained by solving the following problem 
	\begin{subequations}
		\label{poww}
		\begin{align}
			\text{(P2-3.5) :}~
			\mathop{\min}\limits_{\mathbf{c}}~&~\sum\limits_{k\in\mathcal{K}}\varrho_k\epsilon_ke_k
			\\
			\text {s.t.}~&~\eqref{pow1},
		\end{align}
	\end{subequations}
	which can be solved by applying the first-order optimality condition and then determining the dual variable via the bisection method \cite{ming0}.
		\subsubsection{Outer-loop for addressing (P2-2)}
	In the outer loop of PDD framework, the dual variables are updated as
	\begin{align}
		&\mathbf{t}_k \gets  \mathbf{t}_k + \frac{1}{\mu} (\mathbf{w}_k - \mathbf{t}_k), ~\bar{\mathbf{t}} \gets  \bar{\mathbf{t}} + \frac{1}{\mu} (\mathbf{v} - \overline{\mathbf{v}}). \label{dual1}
	\end{align}
	
	Algorithm 1 summarizes the proposed polarforming optimization, which converges \cite{38} with complexity \(\mathcal{O}(I_{\mathrm{out}}I_{\mathrm{in}}KN^2)\), where \(I_{\mathrm{out}}\) and \(I_{\mathrm{in}}\) are the outer and inner iteration numbers required for convergence, respectively.
			\begin{algorithm}[h!]
		\caption{Proposed Polarforming Optimization Algorithm for Solving Problem (P2)}
		\label{alg:PDD}
		\begin{algorithmic}[1]
			\STATE \textbf{Input:} $\mathbf{u}$ and $\zeta$.
			\STATE Initialize $\{\mathbf{w}_k\}$, $\mathbf{v}$, and $\{\mathbf{c}_k\}$, set $i_{\text{out}} = 0$, $\epsilon_{\text{in}} > 0$, $\epsilon_{\text{out}} > 0$, and $ \varpi < 1$.
			\REPEAT
			\STATE Set the inner iteration index $i_{\text{in}} = 0$.
			\REPEAT
			\STATE Update \(\{\mathbf{w}_k\}\), \(\{\bar{\mathbf{w}}_k\}\), \(\mathbf{v}\), \(\bar{\mathbf{v}}\), \(\{\xi_k\}\), \(\{\epsilon_k\}\), and \(\{\mathbf{c}_k\}\) successively.
			\STATE Update the inner iteration index: $i_{\text{in}} \gets i_{\text{in}} + 1$.
			\UNTIL Relative reduction in \eqref{hl9} is below threshold $\epsilon_{\text{in}}$.
			\STATE Update the dual variables by \eqref{dual1} and update $\mu \gets \varpi\mu$.
			\STATE $i_{\text{out}} \gets i_{\text{out}} + 1$.
			\UNTIL Both $\|\mathbf{v} - \bar{\mathbf{v}}\|_\infty$ and $\|\mathbf{w} - \bar{\mathbf{w}}\|_\infty$ fall below $\epsilon_{\text{out}}$, where $\bar{\mathbf{w}}=\{\bar{\mathbf{w}}_k\}$.
			\STATE \textbf{Output:} $\{\mathbf{c}_k\}$,  $\{\mathbf{w}_k\}$, and $\mathbf{v}$.
		\end{algorithmic}
	\end{algorithm}
	%\vspace{-5pt}
	\subsection{Rotation Optimization}
	%\vspace{-3pt}
	In this paper, we assume that the statistical
	CSI of users can be obtained through, e.g., historical data on channel measurements, such that we can focus on the optimization of 
	the BS antenna rotation $\mathbf{u}$ based on the statistical CSI. First, \(L\) channel samples $\hat{\mathcal{H}}^l\!=\!\{\mathbf{h}_1^l,\cdots,\mathbf{h}_K^l\}$ are randomly generated according to the statistical CSI. Problem (P1) is thus recast as 
	\begin{subequations}
		\vspace{-0.38cm}
		\label{lll}
		\begin{align}
			\text{(P3) :}~\mathop{\max}\limits_{\mathbf{u}}~&~
			\sum\limits_{k\in\mathcal{K}}\varrho_k\frac{1}{L}  \sum_{l=1}^{L} R_k\left(
			\mathbf{u}, \mathbf{w}_k,\mathbf{v},\mathbf{c} ,\hat{\mathcal{H}}^l\right) \\
			\text {s.t.}~&~\eqref{M1}.
		\end{align}
	\end{subequations}

	Note that problem (P3) is non-convex due to its non-concave objective function, which makes it challenging to solve. Inspired by the low-complexity particle swarm optimization (PSO) algorithm \cite{con}, we propose a PSO-based scheme for optimizing the rotation parameters. Specifically, we assume that \( S \) particles are used to explore the search space, and the position and velocity of a particle are represented as \(\mathbf{s}\!=\!\mathbf{u}\) and \(\mathbf{d}\), respectively.

  The average sum rate in (P3) serves as the PSO fitness function, which is given by 
	\begin{align}
		J(\mathbf{s}) = \sum\limits_{k\in\mathcal{K}}\varrho_k\frac{1}{L}  \sum_{l=1}^{L} R_k\left(
		\mathbf{u}, \mathbf{w}_k,\mathbf{v},\mathbf{c} ,\hat{\mathcal{H}}^l\right).
		\label{20}
	\end{align}
	
	At iteration \(i\), the \(j\)-th particle updates its position \(\mathbf{s}_j^{(i)}\) based on its velocity \(\mathbf{d}_j^{(i)}\). Let \(\hat{\mathbf{s}}_j\) be the local best position for particle $j$ and \(\mathbf{s}_{\mathrm{g}}\) be the global best position across all particles. The velocity and position updates are, respectively, given by
	\begin{align}
		\mathbf{d}_j^{(i)}&\!=\!\omega\mathbf{d}_j^{(i-1)}\!+\!
		c_1\tau_1(\hat{\mathbf{s}}_{j}^{(i-1)}
		\!-\!\mathbf{s}_j^{(i-1)})\!+\!c_2\tau_2(\mathbf{s}_{\mathrm{g}}^{(i-1)}
		\!-\!\mathbf{s}_j^{(i-1)}),\nonumber\\
		\mathbf{s}_j^{(i)}&=\mathbf{s}_j^{(i-1)}+
		\mathbf{d}_j^{(i)},\label{vb}
	\end{align}
	where \(c_1\) and \(c_2\) are learning factors, \(\tau_1\) and \(\tau_2\) are uniform random variables in \([0,1]\), and \(\omega\) is the inertia weight.
	
	In summary, the two-timescale polarforming and rotation optimization algorithm is outlined in Algorithm 2, in which the complexity of the rotation optimization in Step 1 is \(\mathcal{O}(SLN)\).
			\begin{algorithm}[h!]
		\caption{Two-Timescale Antenna Polarforming and Rotation Optimization}
		\begin{algorithmic}[1]
			\STATE \textbf{Input:} \(L, N, K, S, I_{\text{iter}}\).
			\STATE \textbf{Step 1 (Slow Timescale):} Initialize \(\{\mathbf{s}_j^{(0)}, \mathbf{d}_j^{(0)}\}_{j=1}^S\) and generate \(L\) channel samples.
			\STATE  Compute \eqref{20} using \(\!\{\mathbf{w}_{k,j}^{(0)}, \mathbf{c}_{k,j}^{(0)}\}\!\) and \(\!\{\mathbf{v}_{j}^{(0)}\}\!\) obtained by Alg. 1, and set 
			$\mathbf{s}_{\mathrm{g}}\! =\! \arg \max_{\mathbf{s}} \{\!J^{(0)}(\hat{\mathbf{s}}_1^{(0)}), \dots, J^{(0)}(\hat{\mathbf{s}}_S^{(0)})\!\}$.
			%		\STATE Compute \eqref{20}, where $\{\mathbf{w}_{k,j}^{(0)}, \mathbf{c}_{k,j}^{(0)}\}$ and $\{\mathbf{v}_{j}^{(0)}\}$ are obtained by Algorithm 1.
			%		\STATE 	$
			%		\mathbf{s}_{\mathrm{g}} = \arg \max_{\mathbf{s}} \{J^{(0)}(\hat{\mathbf{s}}_1^{(0)}),  \dots, J^{(0)}(\hat{\mathbf{s}}_S^{(0)})\}$.
			\FOR{\(i = 1\) to \(I_{\text{iter}}\)}
			\STATE Update $\{\mathbf{d}_j^{(i)}\}_{j=1}^S$ and  $\{\mathbf{s}_j^{(i)}\}_{j=1}^S$ via  \eqref{vb}.
			\FOR{\(j = 1\) to \(S\)}
			\STATE Obtain \(\{\mathbf{w}_{k,j}^{(i)}\}\), \(\mathbf{v}_j^{(i)}\), \(\{\mathbf{c}_{k,j}^{(i)}\}\) via Algorithm 1 and evaluate \(J^{(i)}\) according to \eqref{20}.
			\STATE Set \(\hat{\mathbf{s}}_j = \mathbf{s}_j^{(i)}\) if \(J^{(i)} > J^{(i-1)}\); else keep \(\hat{\mathbf{s}}_j = \mathbf{s}_j^{(i-1)}\).
			\ENDFOR
			\STATE Update $
			\mathbf{s}_{\mathrm{g}}= \arg \max_{\hat{\mathbf{s}}} \{J^{(i)}(\hat{\mathbf{s}}_1),  \dots, J^{(i)}(\hat{\mathbf{s}}_S)\}$.
			\ENDFOR
			\STATE \textbf{Output:} \(\mathbf{u}=\mathbf{s}_{\mathrm{g}}\).
			\STATE \textbf{Step 2 (Fast Timescale):} Apply Algorithm 1 using \(\mathbf{u}\) and \(\hat{\mathcal{H}}^l, l\in [1, T_\mathrm{c}]\) to obtain \(\{\mathbf{w}_k\}\), \(\mathbf{v}\) and \(\{\mathbf{c}_k\}\).
		\end{algorithmic}
	\end{algorithm}	
		%\vspace{-3pt}
	\section{Simulation Results}
	%\vspace{-3pt}
	In the simulation, we set \( N = 64 \) and the carrier frequency to 24 GHz. We model the spatial distribution of users over the 3D coverage area using a homogeneous Poisson point process (HPPP) with the average number of users, \(K = 30\). The inter-antenna spacing of the P-6DMA array is \(\frac{\lambda}{2}\). The rotation angles of all users are independently and randomly generated.
	We consider three benchmarks (which are special cases of Algorithm 2): (1) Fixed parameter scheme, where the precoding vector employs maximum-ratio transmission (MRT) and the polarforming vectors’ phase shifts and amplitudes are fixed and randomly generated; (2) Proposed polarforming optimization only, where only the polarforming vectors are optimized using Algorithm 1, while the BS antenna rotations and precoding vectors are the same as in the fixed parameter scheme; (3) Proposed rotation optimization only, where the antenna rotations are optimized via Algorithm 2, with all other parameters being the same as in the fixed parameter scheme.
	\begin{figure}[t!]
			\setlength{\abovecaptionskip}{-3pt}
		\setlength{\belowcaptionskip}{-15pt}
		\centerline{\includegraphics[width=2.72in]{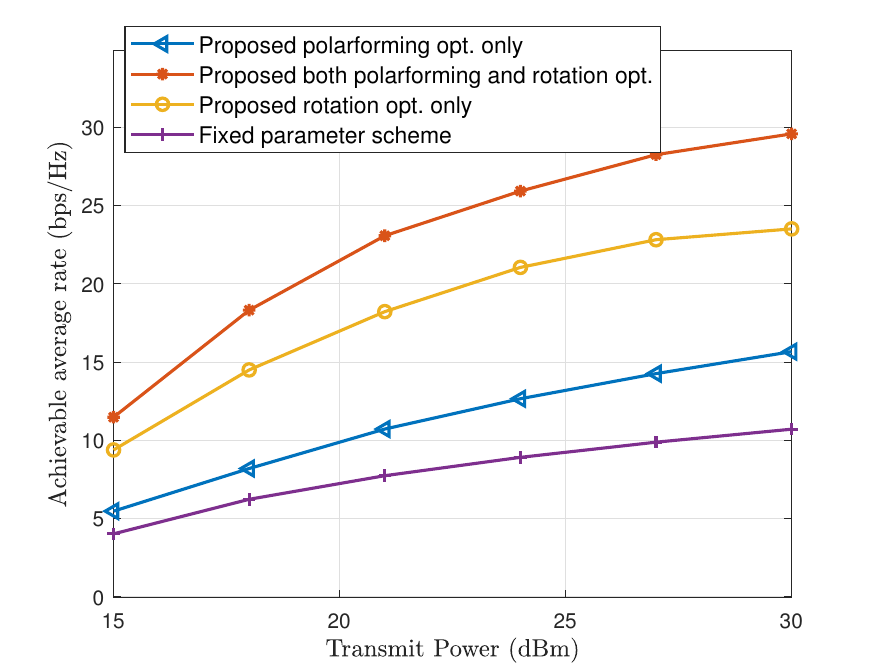}}
		\caption{Achievable average rate  vs. BS transmit power for different schemes.}
		\label{papower}
			\vspace{-0.1cm}
	\end{figure}
	
	In Fig. \ref{papower}, we plot the achievable rates of the different schemes versus (vs.) the BS's transmit power. The results demonstrate that the proposed polarforming optimization only scheme achieves a significant improvement in the average achievable rate over the fixed parameter scheme. This is attributed to the polarforming antenna’s ability to dynamically adjust the antenna polarization of both the users and BS to align the polarization between transmitting and receiving antennas, thereby improving the channel gain. Moreover, it is observed that, with the same transmit power, the rotation optimization-only scheme also consistently achieves higher rates than the fixed parameter scheme. This is because the P-6DMA system with rotation adjustment enjoys more spatial DoFs and can more effectively reshape the antenna radiation pattern in the angular domain to match the users' channel spatial distribution. Furthermore, the joint antenna rotation and polarforming optimization achieves the highest performance gain by exploiting both polarization and spatial diversity.
		
\begin{figure}[t!]
			\setlength{\abovecaptionskip}{-3pt}
			\setlength{\belowcaptionskip}{-15pt}
			\centerline{\includegraphics[width=2.710in]{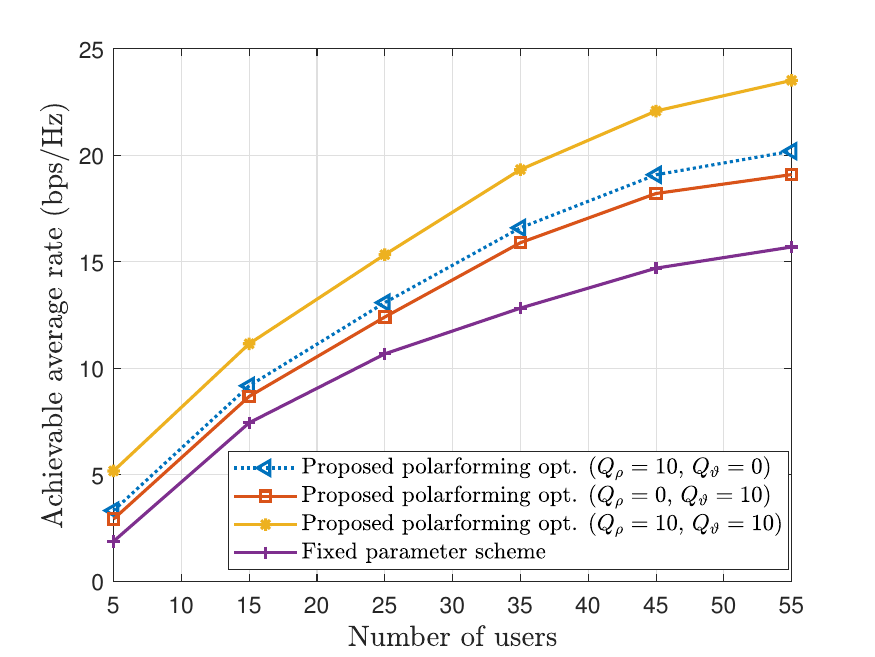}}
			\caption{Achievable rate vs. average number of users for different polarforming quantization levels.}
			\label{bit}
			\vspace{-0.22cm}
		\end{figure}

		In Fig. \ref{bit}, we investigate the achievable sum-rate of the proposed polarforming-optimization-only scheme versus the average number of users, \(K\). We also evaluate the impact
		 of different polarforming amplitude/phase quantization levels on the achievable rate. The number of quantization bits is assumed to be identical at both the BS and users.
		It can be observed that polarforming optimization with joint amplitude and phase control outperforms both phase-only control (i.e., $Q_{\rho}=0$) and amplitude-only control (i.e., $Q_{\vartheta}=0$). Moreover, the proposed schemes with amplitude control achieve better performance than phase-only control. This result suggests that in systems with a large number of users (i.e., interference-limited), it is more beneficial to employ amplitude-phase joint control polarforming rather than amplitude/phase-only polarforming.
		\section{Conclusion}
		In this paper, we proposed a novel P-6DMA to enhance wireless communication performance. First, we modeled the P-6DMA channel in terms of unpolarformed and polarformed components. Then, under the polarforming amplitude and phase constraints, the fast-timescale transceiver polarforming and slow-timescale BS antenna rotation were jointly optimized to maximize the weighted sum rate of users using a low-complexity PDD method. Simulation results have verified the superior performance of our proposed polarforming and rotation optimization methods over various existing benchmark schemes.
		\bibliographystyle{IEEEtran}
		\vspace{-12pt}
		\bibliography{fabs}

\begin{thebibliography}{10}
\providecommand{\url}[1]{#1}
\csname url@samestyle\endcsname
\providecommand{\newblock}{\relax}
\providecommand{\bibinfo}[2]{#2}
\providecommand{\BIBentrySTDinterwordspacing}{\spaceskip=0pt\relax}
\providecommand{\BIBentryALTinterwordstretchfactor}{4}
\providecommand{\BIBentryALTinterwordspacing}{\spaceskip=\fontdimen2\font plus
\BIBentryALTinterwordstretchfactor\fontdimen3\font minus
  \fontdimen4\font\relax}
\providecommand{\BIBforeignlanguage}[2]{{%
\expandafter\ifx\csname l@#1\endcsname\relax
\typeout{** WARNING: IEEEtran.bst: No hyphenation pattern has been}%
\typeout{** loaded for the language `#1'. Using the pattern for}%
\typeout{** the default language instead.}%
\else
\language=\csname l@#1\endcsname
\fi
#2}}
\providecommand{\BIBdecl}{\relax}
\BIBdecl

\bibitem{6gg}
W.~Saad, M.~Bennis, and M.~Chen, ``A vision of {6G} wireless systems:
  Applications, trends, technologies, and open research problems,'' \emph{IEEE
  Netw.}, vol.~34, no.~3, pp. 134--142, May 2020.

\bibitem{larson}
E.~G. Larsson, O.~Edfors, F.~Tufvesson, and T.~L. Marzetta, ``Massive {MIMO}
  for next generation wireless systems,'' \emph{IEEE Commun. Mag.}, vol.~52,
  no.~2, pp. 186--195, Feb. 2014.

\bibitem{wangzhe}
Z.~Wang \emph{et~al.}, ``Extremely large-scale {MIMO}: Fundamentals,
  challenges, solutions, and future directions,'' \emph{IEEE Wireless Commun.},
  vol.~31, no.~3, pp. 117--124, Jun. 2024.

\bibitem{shaos}
X.~Shao, C.~You, W.~Ma, X.~Chen, and R.~Zhang, ``Target sensing with
  intelligent reflecting surface: Architecture and performance,'' \emph{IEEE J.
  Sel. Areas Commun.}, vol.~40, no.~7, pp. 2070--2084, Jul. 2022.

\bibitem{haiquan1}
H.~Lu and Y.~Zeng, ``Communicating with extremely large-scale array/surface:
  Unified modeling and performance analysis,'' \emph{IEEE Trans. Wireless
  Commun.}, vol.~21, no.~6, pp. 4039--4053, Jun. 2022.

\bibitem{duala}
T.~Kim, B.~Clerckx, D.~J. Love, and S.~J. Kim, ``Limited feedback beamforming
  systems for dual-polarized {MIMO} channels,'' \emph{IEEE Trans. Wireless
  Commun.}, vol.~9, no.~11, pp. 3425--3439, Nov. 2010.

\bibitem{pol1}
Y.~He, X.~Cheng, and G.~L. Stuber, ``On polarization channel modeling,''
  \emph{IEEE Wireless Commun.}, vol.~23, no.~1, pp. 80--86, Feb. 2016.

\bibitem{6dmatwc}
X.~Shao, Q.~Jiang, and R.~Zhang, ``{6D} movable antenna based on user
  distribution: Modeling and optimization,'' \emph{IEEE Trans. Wireless
  Commun.}, vol.~24, no.~1, pp. 355--370, Jan. 2025.

\bibitem{poa}
S.-C. Kwon and A.~F. Molisch, ``Capacity maximization with polarization-agile
  antennas in the {MIMO} communication system,'' in \emph{Proc. IEEE Global
  Commun. Conf.}, Dec. 2015, pp. 1--6.

\bibitem{heap}
M.~R. Castellanos and R.~W. Heath, ``Linear polarization optimization for
  wideband {MIMO} systems with reconfigurable arrays,'' \emph{IEEE Trans.
  Wireless Commun.}, vol.~23, no.~3, pp. 2282--2295, Mar. 2024.

\bibitem{38}
Q.~Shi, M.~Razaviyayn, Z.-Q. Luo, and C.~He, ``An iteratively weighted {MMSE}
  approach to distributed sum-utility maximization for a {MIMO} interfering
  broadcast channel,'' \emph{IEEE Trans. Signal Process.}, vol.~59, no.~9, pp.
  4331--4340, Sep. 2011.

\bibitem{ming0}
M.-M. Zhao \emph{et~al.}, ``Intelligent reflecting surface enhanced wireless
  networks: Two-timescale beamforming optimization,'' \emph{IEEE Trans.
  Wireless Commun.}, vol.~20, no.~1, pp. 2--17, Jan. 2021.

\bibitem{con}
W.~Du \emph{et~al.}, ``Network-based heterogeneous particle swarm optimization
  and its application in {UAV} communication coverage,'' \emph{IEEE Trans.
  Emerg. Top. Comput. Intell.}, vol.~4, no.~3, pp. 312--323, Jun. 2020.

\end{thebibliography}
\end{document}